\documentclass[12pt]{article}
\usepackage{amsmath,amsfonts,amssymb,amsthm,amstext,amscd,eucal}
\usepackage[all]{xy}
\usepackage{color}
\usepackage{epsfig}
\usepackage{color}
\usepackage{graphicx}

\makeatletter \@addtoreset{equation}{section}

\makeatletter\renewcommand\section{\@startsection {section}{1}{\z@}%
                                   {-3.5ex \@plus -1ex \@minus -.2ex}
                                   {2.3ex \@plus.2ex}%
                                   {\normalfont\large\bfseries}}
\renewcommand\subsection{\@startsection{subsection}{2}{\z@}%
                                     {-3.25ex\@plus -1ex \@minus -.2ex}%
                                     {1.5ex \@plus .2ex}%
                                     {\normalfont\bfseries}}

\newcommand{\be}{\begin{equation}}
\newcommand{\ee}{\end{equation}}
\newcommand{\bea}{\begin{eqnarray}}
\newcommand{\eea}{\end{eqnarray}}

\newcommand{\bse}{\begin{subequations}}
\newcommand{\ese}{\end{subequations}}
\newcommand{\bi}{\begin{itemize}}
\newcommand{\ei}{\end{itemize}}

\newcommand{\mpl}{M_{\rm pl}}
\newcommand{\mb}{\bar{\mu}}

\begin{document}
\begin{titlepage}

\begin{flushright}\vspace{-3cm}
{\small
{\tt arXiv:1402.2072[hep-th]} \\
\today }\end{flushright}
\vspace{-.5cm}

\begin{center}
\centerline{{\Large{\bf{Black Holes as Beads on Cosmic Strings}}}} \vspace{4mm}

{\large{{\bf Amjad~Ashoorioon\footnote{e-mail: a.ashoorioon@lancaster.ac.uk}$^{,a}$,
Robert B. Mann\footnote{e-mail:
rbmann@uwaterloo.ca}$^{,b}$}}}
\\


\bigskip\medskip
\begin{center}
{$^a$ \it Consortium for Fundamental Physics, Physics Department, Lancaster University,\\ LA1 4YB, United Kingdom}\\
\smallskip

{$^b$ \it Department of Physics, University of Waterloo,\\
Waterloo, Ontario, N2L 3G1, Canada}\\
\end{center}
\vspace{5mm}

\end{center}
\setcounter{footnote}{0}

\date{\today}

\begin{abstract}

We consider the possibility of formation of cosmic strings with black holes as beads. We focus on the simplest setup where two black holes are formed on a long cosmic string. It turns out the in absence of a background magnetic field and for observationally viable values for cosmic string tensions, $\mu<2\times 10^{-7}$, the tension of the strut in between the black holes  has to be less than the  ones that run into infinity. This result does not change if a cosmological constant is present. However if the background magnetic field is turned on, we can have stable setups where the tensions of all cosmic strings are equal. We derive the equilibrium conditions in each of these setups depending on whether the black holes are extremal or non-extremal. We obtain cosmologically acceptable solutions with  solar mass black holes and intragalactic strength cosmic magnatic field.

\end{abstract}

\end{titlepage}
\renewcommand{\baselinestretch}{1.1}

\section{Introduction}

In absence of a background electromagnetic field, two black holes uniformly accelerating from each other are described by the Lorentzian section of the $C-$metric \cite{Kinnersley:1970zw}. Since there is no background field to provide the acceleration, the metric in general has either conical deficits from the black hole to infinity (cosmic strings), or a conical surplus (cosmic rod)  in between the black holes. In the first case, the Euclidean section of the $C-$metric then describes a tunneling process in which a cosmic string is snapped and two black holes are created at the near ends \cite{Hawking:1995zn}.

In presence of a background electromagnetic field, black hole pairs accelerating away from each other are described by the Ernst metric \cite{Ernst1976}. The background magnetic field provides the force that accelerates the (oppositely charged) black holes away from each other. The Euclidean section again describes the process of black hole pair creation in a background magnetic field  \cite{Garfinkle:1993xk}. This is due to the fact that the negative potential energy of the created pair of black holes in the electromagnetic field is balanced by their rest mass energy.  In \cite{Emparan:1995je}, the probability of black hole pair creation with a string running between them was found to be suppressed in comparison with the case in which there is no strings. Pairs of black holes can be also created in presence of cosmological constant and during inflation too \cite{Mann:1995vb,Plebanski:1976gy,Bousso:1996au}.

In all the above papers, it was assumed that only one end of any cosmic string present is attached to  a black holes. We would like to see if configurations in which strings are attached to both ends of the black holes are possible. We are not interested in calculating the instanton solutions  representing the probability of black hole pair creation with strings attached to the  ends; as the probability is often suppressed by the area of the black holes,  cosmic strings attached to them  will reduce the probability further. Instead, we are interested in calculating the classical equilibrium of the configurations irrespective of the mechanism that leading to their formation.

We begin by considering the basic formalism that describes two black holes threaded by a straight cosmic string of infinite length. We find a stable configuration only when the tension of the cosmic string
segments running from the black holes to infinity is larger than the segment between the black holes.
We then examine generalizations of this configuration in the presence of a cosmological constant and of a constant magnetic field, and obtain necessary conditions for stable configurations.  We find  extremal and near-extremal  stable black hole solutions for cosmologically viable values of the intergalactic magnetic field.

\section{Black hole Pairs on Cosmic Strings}

Our starting point is the charged $C$ metric, which is
\be
ds^2=A^{-2}(x-y)^{-2}\left[G(y) dt^2 -G^{-1}(y) dy^2+G^{-1}(x)dx^2+G(x) d\phi^2 \right]
\label{C-metric}\ee
where
\be
G(x)=(1+r_{-} A x)(1-x^2-r_{+} A x^3)
\label{G}
\ee
where
\bea
q^2&=&r_{+} r_{-}\\
m&=&\frac{r_{+}+r_{-}}{2}
\eea
and $A$ is the acceleration parameter. The function $G(x)$ has four roots which we denote by $\xi_1, \xi_2, \xi_3$ and $\xi_4$ in ascending order, i.e.  $\xi_1\leq \xi_2< \xi_3<\xi_4$. To obtain the correct signature we require $\xi_3\leq x\leq \xi_4$ and $-\infty<y\leq x$.  The values $y=\xi_1$, $y=\xi_2$ and $y=\xi_3$ are respectively the inner black hole, the outer black hole and the acceleration horizons. Furthermore, the $x=\xi_3$ and $x=\xi_4$ axes respectively point towards infinity and the other black hole.

\begin{figure}[t]
\begin{center}
\includegraphics[angle=0,
scale=0.50]{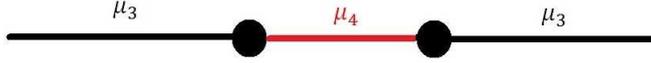}
\caption{Two black holes on a long cosmic string. The tension of the strut between two black holes could be the same or different than the ones that run into infinity.}
\label{Blackholes-on-cosmic-strings}
\end{center}
\end{figure}

We would like to see if one can describe a pair of black holes that have cosmic strings at both poles, running along the axis that connects their centers, see fig.\ref{Blackholes-on-cosmic-strings}.   As we shall demonstrate, such configurations are unstable, unless the tensions of the strings are different.

\subsection{Equal Tension for the Strings}

We consider first the case in which the tension of the strut between the two black holes is the same as the string segments that run to infinity.

 The deficit angles  about the axes $x=\xi_3$ and $x=\xi_4$ are respectively related to the cosmic strings' mass per unit length $\mu_3$ and $\mu_4$ through the following relations
 \be
 \mu_i=\frac{\delta_i}{8\pi},\qquad i=3,4.
 \ee
 Let us find the periodicity $\Delta\phi$ at $x=\xi_4$ which corresponds to the deficit angle $\delta_4$. We require the $(x,\phi)$ section of the $C$-metric near $x=\xi_4$  to have the form $dr^2+r^2 d\Phi^2$. The coordinate $r$ can be found in terms of $x$ through the following relation
 \be
 r\equiv \int \frac{dx}{\sqrt{G(x)}}
 \ee
 Around $x=\xi_4$, $G(x)=(x-\xi_4) G^{\prime}(\xi_4)$, which brings $r$ to the form
 \be
 r=\frac{2\sqrt{x-\xi_4}}{\sqrt{\left|G^{\prime}(\xi_4)\right|}}
 \ee
 or equivalently
 \be
 x-\xi_4=\frac{r^2 G^{\prime}(\xi_4)}{4}
 \ee
 This brings the angular part of the $C$-metric to the cylindrical form, $r^2 d\Phi^2$, around $x=\xi_4$ if we define
 \be
 \Phi=\frac{\left|G^{\prime}(\xi_4)\right|\phi}{2}
 \ee
 Now if we assume that there is a deficit angle at $x=\xi_4$ due to the existence of the strut, {\it i.e.} $\left.\Delta\Phi\right|_{x=\xi_4}=2\pi-\delta_4 $, the $\phi$ periodicity becomes
 \be
 \Delta\phi=\frac{2 (2\pi-\delta_4)}{\left|G^{\prime}(\xi_4)\right|}
 \ee
 Let us now use this to fix the deficit angle at $x=\xi_3$ to be equal to $\delta_3$. Similarly the periodicity $\left.\Delta\Phi\right|_{x=\xi_3}=2\pi-\delta_3$ can be worked out to be
 \be
 \left.\Delta\Phi\right|_{x=\xi_3}=(2\pi-\delta_4)\left|\frac{G^{\prime}(\xi_3)}{G^{\prime}(\xi_4)}\right|.
 \ee
 The above tells us how the tension of the strut and the strings should be related
 \be
 4\mu_3=1-(1-4\mu_4)\left|\frac{G^{\prime}(\xi_3)}{G^{\prime}(\xi_4)}\right|\label{tension-condition}.
 \ee
 Assuming the strut and the strings have the same tension, $\frac{\mu_3}{\mu_4}=1$, we obtain
 \be
4\mu_3\left(1-\left|\frac{G^{\prime}(\xi_3)}{G^{\prime}(\xi_4)}\right|\right)=1-\left|\frac{G^{\prime}(\xi_3)}{G^{\prime}(\xi_4)}\right|.\label{equal-tension-condition}
 \ee
 Note that
 \bea
 G'(\xi_4)&=&(\xi_4-\xi_1)(\xi_4-\xi_2)(\xi_4-\xi_3)>0,\nonumber\\
 G'(\xi_3)&=&(\xi_3-\xi_1)(\xi_3-\xi_2)(\xi_3-\xi_4)<0.\label{G-sign}
 \eea
Let us define
\be\label{sig}
\sigma \equiv \left|\frac{G^{\prime}(\xi_3)}{G^{\prime}(\xi_4)}\right|\leq 1,
\ee
where the upper bound follows from $\xi_1 \leq \xi_2 < \xi_3 < \xi_4$
 There are two ways the condition \eqref{equal-tension-condition} can hold.
 \begin{itemize}
 \item $\sigma \neq 1$: In this case $\mu_3=\mu_4=1/4$. The corresponding deficit angle for this value of cosmic strings tensions is $2\pi$, which is unphysical as in this case the whole space-time is removed.

 \item $\sigma= 1$ and $\mu_3=\mu_4$: The relation $\left|G'(\xi_4)\right|=\left|G'(\xi_3)\right|$ becomes
 \be
 G'(\xi_4)=-G'(\xi_3)\label{G3G4},
 \ee
 Eq. (\ref{G3G4})  will lead to one of the following situations, depending on whether the black holes are extremal or non-extremal:
 \begin{itemize}
 \item \textit{Extremal black holes}: This case corresponds to having $\xi_1=\xi_2$. Eq. \eqref{G3G4} will then imply $\xi_4=\xi_3$, contradicting our assumption about the parameters.

\item \textit{Non-extremal black holes}: In this case $\xi_1<\xi_2$. To have a positive definite Euclidean metric, which is obtained by setting $t=i\tau$ in \eqref{C-metric}, $y$ should be constrained as $\xi_2\leq y\leq \xi_3$. To remove potential singularities at $y=\xi_2$ and $y=\xi_3$, one can fix the $\tau$ period at $y=\xi_3$ to be
    \be
    \Delta \tau=\beta=\frac{4\pi}{G'(\xi_3)},
    \ee
and demand that the surface gravity at $\xi_2$ is the negative of it, {\it i.e.} two horizons having the same temperature \footnote{For the extremal black holes, $\xi_2=\xi_1$, the spatial distance from any other point to $\xi_2$, will be infinite and thus $y=\xi_2$ is not part of the Euclidean section.}\footnote{ Equality of temperatures  at two horizons ensures that there is no conical singularity at outer horizon when extended to the Euclidean section. Since we do not assume that the configuration is obtained from an instantonic transition, one may think that this condition could be relaxed. Assuming that $T_{y=\xi_2}=g T_{y=\xi_3}$ corresponds to replacing eq. \eqref{temp-equality} by
\begin{equation}
G'(\xi_2)=- g G'(\xi_3)\nonumber.
\end{equation}
or equivalently
\be
\xi_2-\xi_1=g(\xi_4-\xi_3)\nonumber.
\ee
Solving the above equation along with \eqref{G3G4}, one can obtain solutions for $\xi_3$ and $\xi_4$. The legitimate pair of solutions for $\xi_4$ and $\xi_3$ are
\bea
\xi_4&=&\frac{\xi_1 + g \xi_1 - \xi_2 + g \xi_2}{2 g}\nonumber\\
\xi_3&=&\frac{-\xi_1 + g \xi_1 + \xi_2 + g \xi_2}{2 g}\nonumber,
\eea
which satisfy the condition $\xi_4>\xi_3$ only if
\be
g<0\nonumber,
\ee
which is unacceptable. Thus the temperature at $y=\xi_2$ and $y=\xi_3$ should be the same.}.
\be\label{temp-equality}
G'(\xi_2)=- G'(\xi_3).
\ee
The last equation is satisfied if
\be
\xi_2-\xi_1=\xi_4-\xi_3. \label{Non-ext-cond}
\ee
Eqs. \eqref{Non-ext-cond} and \eqref{G3G4} together yield $\xi_4=\xi_3$, which again contradicts our assumption about the parameters.
\end{itemize}
 \end{itemize}

Consequently the strut and the string segments both cannot have the same tension. This can be intuitively understood as follows:
the strut  tends to  pull the black holes inward (shortening the strut), whereas   the two string segments tend to pull the
 black holes outward (lengthening the strut).   However the black holes are also attracted towards each other. If the tension of the
 strut and the string segments are the same, the forces exerted by them on the black holes will cancel each other, and the net force remaining is the attraction between two black holes. This will destabilize the configuration, causing it to collapse.

\subsection{Different Tension for the Strings}

If the tensions of two strings are different, $\frac{\mu_3}{\mu_4}=\alpha$, where $\alpha\neq1$, one obtains the following relation using eq. \eqref{tension-condition} ,
\be
\mu_4=\frac{1-\sigma}{4(\alpha-\sigma)}
\label{mu4}
\ee
where $\alpha > \sigma$ must hold so that \eqref{tension-condition} is satisfied. We assume that both $\mu_3$ and $\mu_4$ are both smaller than  the observational upper bound  $\bar{\mu}=2\times 10^{-7}$ \cite{Seljak:2006bg}.

We can constrain  $\sigma|$  using eq.\eqref{mu4}, assuming  that the biggest of the two tensions is smaller than $\bar{\mu}$.  There are two solutions depending on whether $\alpha>1$ or $\alpha<1$
\bi
\item $\alpha<1$: In this case   having the biggest of two tensions smaller than the observational bound, \textit{ i.e.} $\mu_4<\bar{\mu}$,
implies from \eqref{mu4}
\be
\sigma >  \frac{1-4\alpha\bar{\mu} }{1-4\mb}  >1    \label{ext-Gps}
\ee
 which is not possible since $\sigma<1$ from \eqref{sig}.

\item $\alpha>1$: In this case  imposing  the  observational constraint on the largest tension, {\it i.e.} $\mu_3<\bar{\mu}$, yields
from \eqref{mu4}
\be
\frac{\alpha-4\alpha \bar{\mu}}{\alpha-4\bar{\mu}}<
\sigma < 1\label{ext-Gps-a-bigger-one},
\ee
\ei

Note that these conditions hold regardless of whether or not the black holes are extremal.
We conclude that the configuration is stable only when the tension of the cosmic strings running from the black holes to infinity is larger than the strut in between the black holes.

This demonstrates that  one can have black holes as beads on infinitely long cosmic strings, if the tension of the strut in between the black holes is less than the string segments that run to infinity. To have the tensions of the cosmic strings and the strut  both be below the observational limit, the parameters must satisfy the constraints given by \eqref{ext-Gps-a-bigger-one}.

\section{Black Holes on Cosmic Strings in Presence of Cosmological Constant}

As demonstrated above, solutions with equal tensions for the strings and strut cannot be constructed with the $C-$metric (\ref{C-metric}). The presence of a cosmological constant or background magnetic field provides an extra force that might be expected to stabilize such configurations. In the following sections we examine these two possibilities.

In presence of a cosmological constant, $\Lambda$,  the $C-$metric is modified as follows:
\be
ds^2=\frac{1}{A^2 (x-y)^2}\left[H(y)dt^2-H^{-1}(y)dy^2+G^{-1}(x) dx^2+G(x)d\phi^2\right],
\ee
where
\be
G(x)=1-x^2 (1+r_{+}Ax)(1+r_{-}Ax)-\frac{\Lambda}{3A^2},
\ee
and
\be
H(y)=1-y^2 (1+r_{+}Ay)(1+r_{-}Ay),
\ee
The gauge field is
\be
A_{\phi}=q(x-\xi_3)
\ee
and we again only consider the magnetically charged case. The roots of $G(x)$ and $H(y)$ are respectively denoted by $(\xi_1, \xi_2, \xi_3, \xi_4)$ and $(\eta_1,\eta_2,\eta_3,\eta_4)$, each in ascending order. As with the C-metric, $y=\eta_1$, $y=\eta_2$ and $y=\eta_3$ are respectively the inner black hole, the outer black hole and the acceleration horizons.

One can again envisage two cases where the tensions of the cosmic strings are equal or different, and the black holes can be extremal or non-extremal. Note that equations \eqref{equal-tension-condition}, \eqref{G-sign}, and \eqref{sig} hold in this case as well.  Hence the analysis proceeds similarly to the $\Lambda=0$ case.

\subsection{Equal tension for Cosmic Strings}

Assuming the strut and the strings have the same tension, $\frac{\mu_3}{\mu_4}=1$, eq. \eqref{equal-tension-condition} should hold to have a stable solution. Thus, as in the case in which there is no cosmological constant, the only stable configuration with equal cosmic string tension is either obtained by assuming that $\mu_3=\mu_4=\frac{1}{4}$ (leading to the unphysical deficit angle $\delta=2\pi$) or by
 setting $G'(\xi_4)=-G'(\xi_3)$.
  This latter constraint yields
 \be
(\xi_4-\xi_1)(\xi_4-\xi_2)=(\xi_3-\xi_1)(\xi_3-\xi_2)\label{eq-tension-Lambda-Ext}
\ee
for both the extremal and non-extremal cases. It is straightforward to show that
\eqref{eq-tension-Lambda-Ext} cannot be solved if $\xi_1<\xi_2<\xi_3<\xi_4$, since the left-hand side of this equation
is larger in magnitude than the right-hand side.

Since the equal-tension case is ruled out, we now consider the case of unequal tensions.

\subsection{Unequal tensions for Cosmic Strings}

Assuming $\alpha\equiv \frac{\mu_3}{\mu_4}$, one can solve eq. \eqref{tension-condition} to obtain $\mu_4$ as in eq. \eqref{mu4}.
The analysis is identical to the $\Lambda=0$ case.  Hence we conclude that for nonzero cosmological constant
 the parameters must satisfy the constraints given by \eqref{ext-Gps-a-bigger-one} if the tensions of the cosmic strings and the strut are both below the observational limit.

\section{Black Holes on a Cosmic Strings in Presence of Background Magnetic Field}

In the presence of a background magnetic field the $C-$metric must be replaced by the Ernst metric which describes the motion of charged black holes accelerating in opposite directions \cite{Ernst1976}
\bea
ds^2&=&\frac{\Gamma^2}{A^2 (x-y)^2}\left[G(y)dt^2 -G^{-1}(y)dy^2+G^{-1}(x) dx^2\right]+\frac{G(x)}{\Gamma^2 A^2 (x-y)^2} d\phi^2,\\
A_{\phi}&=&-\frac{2}{B\Gamma}(1+\frac{1}{2}B q x)+k.
\eea
$G(x)$ with four roots $\xi_4>\xi_3>\xi_2>\xi_1$ is given as above and $\Gamma$ is
\be
\Gamma=(1+\frac{1}{2}B q x)^2+\frac{B^2}{4A^2(x-y)^2}G(x).
\ee
Note that $\Gamma(\xi)\equiv \Gamma(\xi_j,y) = (1+\frac{1}{2}B q \xi_j)^2$ is independent of $y$.

$k$ is chosen to confine the Dirac string singularities to the $x=\xi_4$ axis. The metric  reduces to the Melvin metric \cite{Melvin:1963qx},
\be
ds^2=(1+\frac{1}{4}B^2 \rho^2)^2 (-dt^2+dz^2+d\rho^2)+\frac{\rho^2 d\phi^2}{(1+\frac{1}{4}B^2 \rho^2)^2},
\ee
\be
A_{\phi}=\frac{\rho^2 B}{2(1+\frac{1}{4}B^2 \rho^2)},
\ee
at spatial infinity, $x,y\rightarrow\xi_3$.

We are again interested in the situation where there are conical singularities at the axis running to infinity and in between the black holes. As with the $C-$metric, we compute the relation between the deficit angles, or equivalently the tensions, which turns out to be
\be
4\mu_3=1-(1-4\mu_4)\left|\frac{G'(\xi_3)}{G'(\xi_4)}\right|\frac{\Gamma(\xi_4)^2}{\Gamma(\xi_3)^2}.\label{tensions-relation-magnetic-field}
\ee
Unlike the previous cases, the quantity $\frac{\Gamma(\xi_4)^2}{\Gamma(\xi_3)^2}$ can be larger than unity, and so new possibilities emerge for
placing black holes on the cosmic string.

 Setting
\be\label{gam}
\gamma =\left|\frac{G'(\xi_3)}{G'(\xi_4)}\right|\frac{\Gamma(\xi_4)^2}{\Gamma(\xi_3)^2}
\ee
we find that \eqref{mu4} becomes
\be
\mu_4=\frac{1-\gamma}{4(\alpha-\gamma)}
\label{mu4b}
\ee
and we can proceed with the analysis as before, replacing $\sigma\to \gamma$ where  $\gamma>0$.

Below, we will consider the cases in which strings tensions are equal and different separately.

\subsection{Equal String Tensions}

Similar to the no background magnetic field case, requiring equal tensions $\mu_3=\mu_4=\mu$ for the strings lead to the following two cases
\bi
\item $\gamma\neq 1$: In this case the tension of all string parts should be equal to $\mu=\frac{1}{4}$. This case corresponds to the unphysical deficit angle of $2\pi$ which is unattainable.

\item $\gamma= 1$: In this case, $\mu$ can take any value, even though we are interested in observationally relevant values which constrain $\mu<\bar{\mu}=2\times 10^{-7}$. Noting \eqref{mu4b}, this condition could be equivalently written as
\be
-\frac{G'(\xi_4)}{\Gamma^2(\xi_4)}=\frac{G'(\xi_3)}{\Gamma^2(\xi_3)}\label{Magnetic-field-con}
\ee
\ei
We therefore conclude that two black holes can be placed on a cosmic string in a constant magnetic field provided
\eqref{Magnetic-field-con} is satisfied.  In the limit $B\to 0$  eq. \eqref{sig} is recovered, indicating that there are no solutions with equal tension unless the magnetic field is nonzero.  There is also a solution
in the large-$B$ limit that recovers eq. \eqref{sig} for small accelerations as we shall see. Hence
equation (\ref{Magnetic-field-con}) in general
 results in two possible solutions
\be\label{Bsoln}
B = \pm 2\frac{ \lambda \mp 1 }{q\left(\xi_4 \mp  \lambda \xi_3\right) }
\ee
for the magnetic field,
where
\be\label{orig-lambda}
\lambda = \left[\frac{(\xi_4-\xi_1)(\xi_4-\xi_2)}{(\xi_3-\xi_1)(\xi_3-\xi_2)}\right]^{1/4} > 1.
\ee

In the extremal case, which corresponds to $\xi_1=\xi_2$, $\lambda$ simplifies to
\be
\lambda_{\rm ext}=\left[\frac{\xi_4-\xi_1}{\xi_3-\xi_1}\right]^{1/2},
\ee
whereas in the non-extremal case $\lambda$ simplifies to
\be
\lambda_{\rm Next.}=\left[\frac{\xi_4-\xi_1}{\xi_3-\xi_2}\right]^{1/4}
\ee
upon using eq. \eqref{Non-ext-cond} since the black holes are at the same temperature.

\subsection{Unequal Tension for Cosmic Strings}

Proceeding as before we consider the $\alpha <1$ and $\alpha >1$ cases separately.

\bi
\item $\alpha<1$: In this case   having the biggest of two tensions smaller than the observational bound, \textit{ i.e.} $\mu_4<\bar{\mu}$,
implies from \eqref{mu4b}
\be
\gamma = \left|\frac{G'(\xi_3)}{G'(\xi_4)}\right|\frac{\Gamma(\xi_4)^2}{\Gamma(\xi_3)^2} >  \frac{1-4\alpha\bar{\mu} }{1-4\mb}  >1    \label{ext-Gpsb}
\ee
 which is possible provided $\gamma > 1$.

\item $\alpha>1$: In this case  imposing  the  observational constraint on the largest tension, {\it i.e.} $\mu_3<\bar{\mu}$, yields
from \eqref{mu4b}
\be
\frac{\alpha-4\alpha \bar{\mu}}{\alpha-4\bar{\mu}}<\left|\frac{G'(\xi_3)}{G'(\xi_4)}\right|\frac{\Gamma(\xi_4)^2}{\Gamma(\xi_3)^2}=
\gamma < 1\label{ext-Gps-a-bigger-oneb},
\ee
\ei
In principle it should be possible to find legitimate values for the parameters such that the above constraints are satisfied in both two cases.
In the next section we obtain such values for a few examples.

\section{Examples}

The above semi-classical treatment is only valid for black holes whose mass is larger than $\mpl$. For illustrative purposes, in what follows we set the mass of the black hole, $m \sim 10 \mpl$. We also assume that the magnetic charge of the black hole takes a value in between $-10\leq q\leq 10$. We then have to adjust the acceleration parameter, $A$, such that we have four distinct roots and the obtained values for the tensions are below the observational limit.

 Let us first focus on the case where there is no cosmological constant or background magnetic field and the black holes are non-extremal, {\it i.e.} $m\neq |q|$. For definiteness let us assume that $q=5$. The tensions of the strut and the strings that run to infinity should be different. Let us assume that $\alpha\equiv \mu_3/\mu_4=2$. For such values of $m$, $q$ and $\alpha$ we find from eq. \eqref{ext-Gps-a-bigger-one}  that $A$ has to be equal or smaller than $10^{-8}$ to have $\mu_3\leq \bar{\mu}$. We obtain  $\xi_1$ and $\xi_2$   by solving the $G(x)=0$ numerically and find they are respectively equal to $-7.46\times 10^7$ and $\simeq -5.36\times 10^6$, and so the black holes are non-extremal (the other two roots are respectively equal to $\xi_3\simeq-1$ and $\xi_4\simeq 0.999$). We find that by varying $q$, in the allowed range, $-10< q < 10$, the maximum value for $A$ does not change significantly.

One can come up with an example of extremal black holes too. For example if we assume $m=q=5$ and $A=2.01\times 10^{-8}$, $\xi_1$ and $\xi_2$ turn out be equal to 25 significant figures: $\xi_1\simeq \xi_2\simeq -9.95 \times 10^{6}$ (the other two roots are respectively equal to $\xi_3\simeq-1$ and $\xi_4\simeq 0.999$). In this case we find that $\alpha\gtrsim 2.01$, to have $\mu_3<\bar{\mu}$.

We can repeat the calculations in the presence of a cosmological constant. Setting the cosmological constant equal to today's value, $\Lambda\simeq10^{-122}$, we find that the roots do not change that much, and as before $A$ needs to be tuned to be equal or smaller than $\lesssim 10^{-8}$ in order to have $\alpha\gtrsim 1.999$. During the inflationary era the value of cosmological constant  is larger. For $m=q=10$ and $A=10^{-8}$, the largest value we obtained for the cosmological constant for which four real roots were available was $\Lambda=2\times 10^{-16}$. In order to have the biggest of two tensions smaller than the observational limit we must set $\alpha\gtrsim 1.40$.

For the values of the parameters we obtained above for extremal black holes, setting the cosmological constant to its present value
($\Lambda\simeq10^{-122}$) does not alter the value of the roots that much and thus one is left with an almost extremal black hole. To have the tensions below the observational limit, we find that $\alpha \gtrsim  1.999$. For the same set of parameters it is possible to increase $\Lambda$ up to $\simeq 10^{-15}$ before two of the roots become imaginary. Even for $\Lambda=10^{-15}$,  the black holes are almost extremal. In order to have the tensions below the observational limit the ratio of tensions we find $\alpha>1.28$.

When the magnetic field is nonzero
we obtain a small-$B$ and a large-$B$ solution consistent with our discussion in the previous section. We shall see below that this occurs for every example we consider.

For the former set of parameters above, where we had non-extremal black holes, one can assume that the tension of strings are equal if the background magnetic field is turned on.
 For the non-extremal example above,
we find
\be
B\simeq 2\times 10^{-8}\qquad \mathrm{or} \qquad B\simeq 2.79\times 10^6
\ee
 for the magnetic field.

Next suppose that the tensions of the strings are unequal. For the former set of black hole parameters, where the black holes were non-extremal, assuming that $\alpha=1/2$, we obtain the following  intervals for $B$
\be
4\times 10^{-8}<B<2.06\times 10^{6}
\ee
such that the tensions remain below the observational bound.  Retaining the same set of parameters,
if $\alpha=2$,  we find
\be
B<2\times 10^{-8} \qquad \mathrm{or} \qquad 2.79\times 10^{6}<B<4.28\times 10^{6}
\ee

For the latter set of parameters, where we have extremal black holes, one can again assume that the tensions of cosmic strings are all equal and find the corresponding values of background magnetic field. Numerically solving the equation  $B$ has to be either equal to
\be
B=1\times 10^{-7}  \qquad\mathrm{or} \qquad B=1.99\times 10^{7}.
\ee
Again for negative but equally charged black holes, it is not possible to satisfy condition \eqref{Magnetic-field-con}.
If we assume that the tensions are unequal and $\alpha=1/2$, we obtain a observationally viable value for largest string tension,  $\mu_4$, provided
\be
4.01\times 10^{-8}<B< 2.66\times10^{6}.
\ee
If we set $\alpha=2$, the constraint $\mu_3\leq \bar{\mu}$, limits $B$ as follows
\be
9.87\times 10^{-11}<B<2.01\times 10^{-8} \qquad \mathrm{or}\qquad 3.99\times10^{6}<B<7.98\times 10^6.
\ee
The above value of the magnetic fields are in natural units. Recovering the factors of $\hbar$, $c$, $G$ and $\epsilon_0$, the magnitude of magnetic field is in units of $\frac{c^{5/2}}{\hbar^{1/2} G {\epsilon_0}^{1/2}}$ which is approximately $7.63\times 10^{53}$ T, or $7.63\times 10^{57}$ G. Thus, the values of obtained magnetic fields are much larger than the intragalactic magnetic fields \cite{Ashoorioon:2004rs} which is about $\lesssim 10^{-6}$ G.

One may wonder if it is possible to have black holes on cosmic strings with cosmologically viable values for the magnetic field. It is straightforward to verify that for the extremal black holes with $m=q=10$, taking $A=10^{-68}$, one obtains the following values for the magnetic fields
\be
B\simeq  10^{-68} \qquad \mathrm{or} \qquad B\simeq 2\times 10^{66},
\ee
in natural units, assuming that the tension of the strut and cosmic strings are the same. Converting the units to the SI units, the first solution corresponds to $7.63\times 10^{-11}$ T or $7.63\times 10^{-7}$ G, which is in the range of observationally interesting values for   cosmic magnetic fields. One can also assume the masses of the extremal black holes are close to the mass of the Sun, $M_{\odot}\simeq 10^{38} \mpl$. Assuming $A=10^{-68}$, we obtain
\be
B\simeq  10^{-68} \qquad \mathrm{or} \qquad B\simeq 2\times 10^{-8},
\ee
where the first solution is in the range of intragalactic magnetic fields.

Since both large-$B$ and small-$B$ solutions to \eqref{Bsoln} exist, we expect that for small
$A$ the field is respectively proportional and inversely proportional to $A$.  For semiclassical black holes, like the ones above, with $r_{-}A \leq 1$ and $ r_{+}A\ll 1$, the $\xi_i$ roots are as follows
\bea
\xi_1&\simeq& - \frac{1}{r_{-}A}\ ,\qquad \xi_2\simeq -\frac{1}{r_{+}A}+r_{+}A\ ,\\
\xi_3&\simeq& -1-\frac{r_{+}A}{2}\ ,\qquad \xi_4\simeq 1-\frac{r_{+}A}{2}
\eea
Note that to leading order in $A$ both the large-$B$ and small-$B$ limits yield eq. \eqref{sig}.
To first order in $r_{-}A$ and $r_{+}A$, $\lambda$ in eq. \eqref{orig-lambda} is
\be
\lambda\simeq 1+\frac{r_{-}+r_{+}}{2q} A=1+\frac{m}{q} A
\ee
In such a limit, the solutions for the magnetic field \eqref{Bsoln} are
\be
B\simeq  \frac{r_{-}+ r_{+}}{ 2q}A=\frac{m}{q}A \ ,\qquad {\rm or}  \qquad B\simeq
\frac{8}{q(r_{-}+ 3 r_{+})A} \quad .
\ee
For extremal black holes $q=m$, the solutions for magnetic fields become
\be
B\simeq A \ ,\qquad {\rm or}  \qquad B\simeq \frac{2}{m^2 A},
\ee
as expected, explaining the equality of the first solutions and the huge gap between the second ones for different black hole masses above.

\section{Conclusion}

 Using C-metric and its extensions, with the cosmological constant or the magnetic field, we obtained solutions that describe two black holes are on
 a long cosmic string. We showed that, without the magnetic field, the tension of the strut between the black holes has to be less than the tension of the outer strings running to infinity. Only in the presence of a  background magnetic field is it  possible to have black holes on a cosmic string with uniform tension.

Although the parameters must be appropriately adjusted, cosmologically viable solutions with solar mass black holes exist for intragalactic strength cosmic magnetic field strength. It would be interesting so see if more detailed numerical solutions for such black hole beads could be constructed in which the string is represented by a vortex \cite{oai:arXiv.org:gr-qc/9506054} so as to better study the phenomenological prospects of these kinds of configurations.

The equilibrium of the above configuration in presence of a cosmic magnetic field is independent
of the tension of the cosmic string and only relies on the black hole mass and charge
parameters. A large massive black hole can be stabilized by a small tension cosmic string
in the presence of an appropriate magnetic field. Thus the lensing \cite{oai:arXiv.org:astro-ph/9702033} and gravitational waves
emitted from cusps on such strings \cite{oai:arXiv.org:1002.0652} could be in principle be enhanced by the presence of
black holes even if the tension of strings is below the detectability limit of probes. One can
also investigates the effect of such configuration on the two point correlation of the inflaton
\cite{ARXIV:0908.0543}. This gives further motivation to investigate these signatures of the setup in future.

\section*{Acknowledgements}
R.B.M. is supported in part by the Natural Sciences and Engineering Research Council of Canada. A.A. is supported by the Lancaster-Manchester-Sheffield Consortium for Fundamental Physics under STFC grant ST/J000418/1.

\end{document}